\documentclass{naturefig}

\newcommand{\apj}{\textit{Astrophys. J.}}
\newcommand{\apjl}{\textit{Astrophys. J. Lett.}}
\newcommand{\apjs}{\textit{Astrophys. J. Supp.}}
\newcommand{\mnras}{\textit{Mon. Not. R. Astron. Soc.}}

\newcommand{\aap}{\textit{Astron. \& Astrophys.}}
\newcommand{\aaps}{\textit{Astron. \& Astrophys. Supp.}}

\newcommand{\aj}{\textit{Astron. J.}}

\newcommand{\pasj}{\textit{Pub. Astron. Soc. Jap.}}

\newcommand{\jcap}{\textit{JCAP}}

\usepackage{hyperref}
\usepackage{color}
\usepackage{epsfig,psfrag,graphics,verbatim}
\usepackage{graphicx}
\usepackage{amsmath,amssymb}
\usepackage{ulem}
\usepackage{times}
\usepackage{caption}
\usepackage{siunitx}

\title{Evidence for dark matter in the inner Milky Way}

\author{Fabio Iocco$^{1,2}$, Miguel Pato$^{3,4}$ \& Gianfranco Bertone$^5$}

\begin{document}
\spacing{1}

\maketitle

\vspace{-3.3cm}
\begin{flushright}
\small TUM-HEP 973/15
\end{flushright}
\vspace{1.3cm}

\begin{affiliations}
 \item Instituto de F\'isica Te\'orica UAM/CSIC, C/ Nicol\'as Cabrera 13-15, 28049 Cantoblanco, Madrid, Spain;
 \item ICTP South American Institute for Fundamental Research, and Instituto de F\'isica Te\'orica - Universidade Estadual Paulista (UNESP), Rua Dr. Bento Teobaldo Ferraz 271, 01140-070 S\~{a}o Paulo, SP Brazil;
 \item Physik-Department T30d, Technische Universit\"at M\"unchen, James-Franck-Stra\ss{}e, 85748 Garching, Germany;
 \item The Oskar Klein Centre for Cosmoparticle Physics, Department of Physics, Stockholm University, AlbaNova, SE-106 91 Stockholm, Sweden;
 \item GRAPPA Institute, University of Amsterdam, Science Park 904, 1090 GL Amsterdam, The Netherlands.
\end{affiliations}

\begin{abstract}
The ubiquitous presence of dark matter in the universe is today a central tenet in modern cosmology and astrophysics\cite{BertoneBook}. Ranging from the smallest galaxies to the observable universe, the evidence for dark matter is compelling in dwarfs, spiral galaxies, galaxy clusters as well as at cosmological scales. However, it has been historically difficult to pin down the dark matter contribution to the total mass density in the Milky Way, particularly in the innermost regions of the Galaxy and in the solar neighbourhood\cite{JustinRead2014}. Here we present an up-to-date compilation of Milky Way rotation curve measurements\cite{Fich1989,McClure-GriffithsDickey2007,Luna2006,HonmaSofue1997,BrandBlitz1993,Hou2009, FrinchaboyMajewski2008,Durand1998,Pont1997,Battinelli2013,Reid2014}, and compare it with state-of-the-art baryonic mass distribution models\cite{Stanek1997,LopezCorredoira2007,Robin2012,deJong2010,Juric2008,BovyRix2013,Ferriere1998,Moskalenko2002}. We show that current data strongly disfavour baryons as the sole contribution to the galactic mass budget, even inside the solar circle. Our findings demonstrate the existence of dark matter in the inner Galaxy while making no assumptions on its distribution. We anticipate that this result will compel new model-independent constraints on the dark matter local density and profile, thus reducing uncertainties on direct and indirect dark matter searches, and will shed new light on the structure and evolution of the Galaxy. 
\end{abstract}

\par Existing studies of the dark matter density in the inner Galaxy fall into two categories: mass-modelling and local measurements. In {\it mass-modelling}, the distribution of dark matter is assumed to follow a density profile inspired by numerical simulations, typically an analytic fit such as the well-known Navarro-Frenk-White\cite{NFW1996} or Einasto\cite{Merritt2006} profiles, with two or more free parameters whose best-fit values are then determined from dynamical constraints. The statistical error on the dark matter density in the inner Galaxy -- and in particular in the solar neighbourhood -- is in this case very small, of order 10\%\cite{CatenaUllio2010}, but this only reflects the strong assumptions made about the distribution of dark matter. The latter is in fact observationally unknown, and the aforementioned classes of profiles are inspired by simulations without baryons, whose role is not negligible in the inner Galaxy. {\it Local measurements} are instead based on the study of observables in the solar neighbourhood\cite{JustinRead2014}. These methods can be used to assess the evidence for dark matter locally through an estimate of the gravitational potential from the kinematics of stars. However, the value found for the local dark matter density is usually compatible with zero at $\sim$3$\sigma$ unless one makes strong assumptions about the dynamics of the tracer populations.

\par Here we report on a comparison of the observed rotation curve of the Galaxy with that expected from visible matter only. As we shall see, this approach provides an alternative way to constrain additional contributions of matter to the rotation curve, and therefore to infer the existence and abundance of dark matter. Although this has been historically one of the first methods to detect dark matter in external galaxies, it has long been thought to provide weak constraints in the innermost regions of the Milky Way, due to a combination of poor rotation curve data and large uncertainties associated with the distribution of baryons. We show that recent improvements on both fronts permit to obtain a convincing proof of the existence of dark matter inside the solar circle.

\par We start by presenting a new, comprehensive compilation of rotation curve data derived from kinematic tracers of the galactic potential, which considerably improves upon earlier (partial) compilations\cite{Sofue2009,Bhattacharjee2014}. Optimised to galactocentric distances $R=3-20$ kpc, our database includes gas kinematics  (HI terminal velocities\cite{Fich1989,McClure-GriffithsDickey2007}, CO terminal velocities\cite{Luna2006}, HI thickness\cite{HonmaSofue1997}, HII regions\cite{BrandBlitz1993,Hou2009}, giant molecular clouds\cite{Hou2009}), star kinematics (open clusters\cite{FrinchaboyMajewski2008}, planetary nebulae\cite{Durand1998}, classical cepheids\cite{Pont1997}, carbon stars\cite{Battinelli2013}) and masers\cite{Reid2014}. This represents an exhaustive survey of the literature that intentionally excludes objects with kinematic distances only, and those for which asymmetric drift or large random motions are relevant. In total we have compiled 2780 measurements, of which 2174, 506 and 100 from gas kinematics, star kinematics and masers, respectively (see supplementary information). For each measurement, we translate the kinematic data into a constraint on the angular velocity $\omega_c=v_c/R$ and on the galactocentric distance $R$. The upper panel of Fig.~1 shows the rotation curve $v_c(R)$ for the full compilation of data.

\par The contribution of stars and gas to the total mass of the Galaxy has historically been subject to significant uncertainties, in particular towards the innermost regions where its dynamical contribution is most important. Substantial progress has been made recently, and data-based models that encode the three-dimensional morphology of the baryonic distribution have become available in the literature. In order to bracket the uncertainties on the stellar and gas distribution, we consider here all possible combinations of a set of detailed models for the stellar bulge\cite{Stanek1997,LopezCorredoira2007,Robin2012}, stellar disk\cite{deJong2010,Juric2008,BovyRix2013} and gas\cite{Ferriere1998,Moskalenko2002} (see supplementary information). The stellar bulge models encompass alternative density profiles in the inner few kiloparsecs and different configurations of the galactic bar. The stellar disks implemented provide instead the best descriptions of star observations across the Galaxy, including parametrisations with and without separation into thin and thick populations. Finally, the gas is split into its molecular, atomic (cold, warm) and ionised (warm, hot, very hot) components, paying special attention to the localised features in the range $R=10$ pc$-20$ kpc. 

\par The gravitational potential of each model is computed through multipole expansion\cite{BinneyTremaine} (see supplementary information), and the corresponding rotation curve is shown in the lower panel of Fig.~1 with its original normalisation. We calibrated each bulge with the microlensing optical depth measurement towards the central galactic region $\langle \tau \rangle = 2.17^{+0.47}_{-0.38}\times 10^{-6}$\cite{MACHO2005,Iocco2011}, each disk with the constraint on the local surface density $\Sigma_{*}=38\pm 4\textrm{ M}_\odot\textrm{/pc}^2$\cite{BovyRix2013}, and for the gas we adopted a CO-to-H$_2$ conversion factor of $(0.5-3.0)\times 10^{20}\textrm{ cm}^{-2}\textrm{(K km/s)}^{-1}$ for $R>2$ kpc\cite{Ferriere1998,Ackermann2012}. This procedure ensures that all baryonic models comply with the existing observational constraints and moreover it assigns a realistic uncertainty to the contribution of each model to the rotation curve.

\par We assess the evidence for an unseen (dark) component of the gravitational potential of our Galaxy in the form of a discrepancy between the observed rotation curve and that expected from the set of baryonic models described above. We stress that we do not make any assumption about the nature or distribution of dark matter: our analysis therefore provides a model-independent estimate of the amount of dark matter in the Galaxy. For each baryonic model, the two-dimensional chi-square\cite{Fich1989} is computed and used to assess the goodness-of-fit. We have explicitly checked through Monte Carlo calculations that this statistic has an approximate $\chi^2$ distribution for the case at hand. The analysis is restricted to galactocentric distances $R>R_{cut}=2.5$ kpc, below which the orbits of the kinematic tracers are significantly non-circular. We adopt a distance to the galactic centre $R_0=8$ kpc, a local circular velocity $v_0=230$ km/s, and a peculiar solar motion\cite{Schoenrich2010} $\left(U,V,W\right)_{\odot}=(11.10,12.24,7.25)$ km/s.

\par The upper panel of Fig.~2 shows the angular velocity as a function of the galactocentric distance. Observational data are shown with red dots, while the grey band shows the envelope of all baryonic models discussed above, which we interpret here as bracketing the possible contribution of baryons to the rotation curve. The discrepancy between observations and the expected contribution from baryons is evident along the whole range of galactocentric distances above $6-7$ kpc. The residuals are plotted in the middle panel of Fig.~2 for a fiducial baryonic model (the one shown by the solid black line in the upper panel), and they can be readily interpreted as the contribution of an extra component to the Newtonian gravitational potential of our Galaxy. Interestingly, the gravitational potential from a dark matter distribution such as those suggested by numerical simulations (Navarro-Frenk-White or Einasto profiles) smoothly fills the gap without fine tuning.

\par The main conclusion of our analysis is summarised in the bottom panel of Fig.~2, where we plot the $\chi^2$ per degree of freedom for each baryonic model and for all data up to a given distance $R$ (but above $R_{cut}$). The evidence for a dark component rises above 5$\sigma$ (thick red line) well inside the solar circle for all baryonic models. Indeed, whereas the relative discrepancy between observational data and baryonic models is higher at larger galactocentric distances, it is at lower distances that uncertainties are smallest. Hence, the evidence grows swiftly at relatively small distances and saturates at larger distances. We have tested the robustness of our results against variations of $R_0$, $v_0$, peculiar solar motion, binning and data selection as well as against systematics due to spiral arms\cite{BrandBlitz1993}. The results change only mildly for all cases, and the conclusions drawn from Fig.~2 remain unchanged (see supplementary information).

\par The comparison of the Milky Way observed rotation curve with the predictions of a wide array of baryonic models points strongly to the existence of a contribution to the gravitational potential of the Galaxy from an unseen, diffuse component. The statistical evidence is very strong already at small galactocentric distances, and it is robust against uncertainties on galactic morphology and kinematics. Without any assumption about the nature of this dark component of matter, our results open a new avenue for the determination of its distribution inside the Galaxy. This has powerful implications both on studies aimed at understanding the structure and evolution of the Milky Way in a cosmological context, and on direct and indirect dark matter searches, aimed at understanding the very nature of dark matter.

\begin{figure}
\includegraphics[width=1\textwidth]{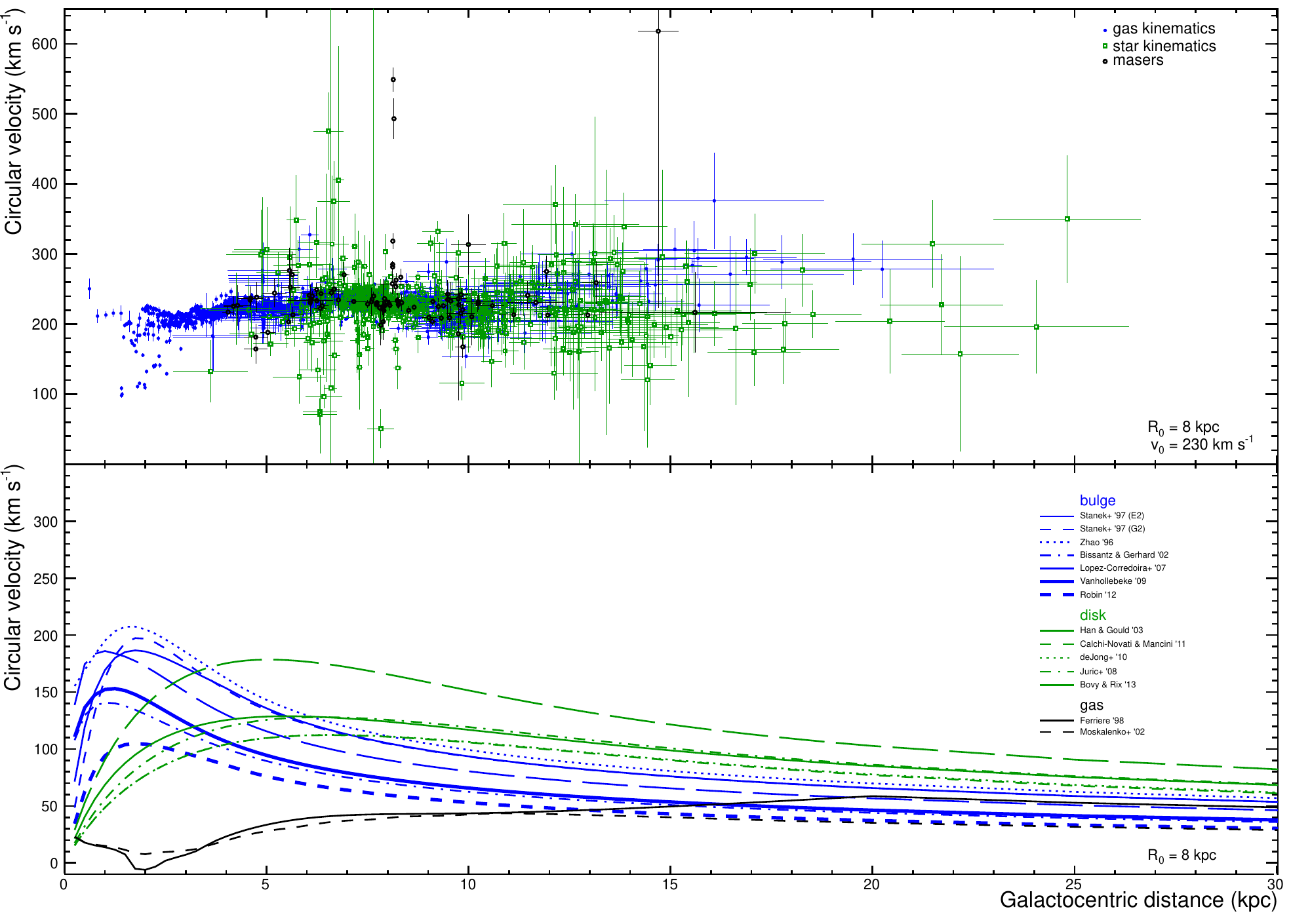}
\caption{The rotation curve of the Milky Way. In the top panel we show our compilation of rotation curve observations as a function of galactocentric distance, including data from gas kinematics (blue dots; HI terminal velocities, CO terminal velocities, HI thickness, HII regions, giant molecular clouds), star kinematics (open green squares; open clusters, planetary nebulae, classical cepheids, carbon stars) and masers (open black circles). Error bars correspond to 1$\sigma$ uncertainties. The bottom panel displays the contribution to the rotation curve as predicted from different models for the stellar bulge (blue), stellar disk (green) and gas (black). We assume a distance to the galactic centre $R_0=8$ kpc in both panels, and a local circular velocity $v_0=230$ km/s in the top panel.}\label{fig:data_models}
\end{figure}

\begin{figure}
\includegraphics[width=1.\textwidth]{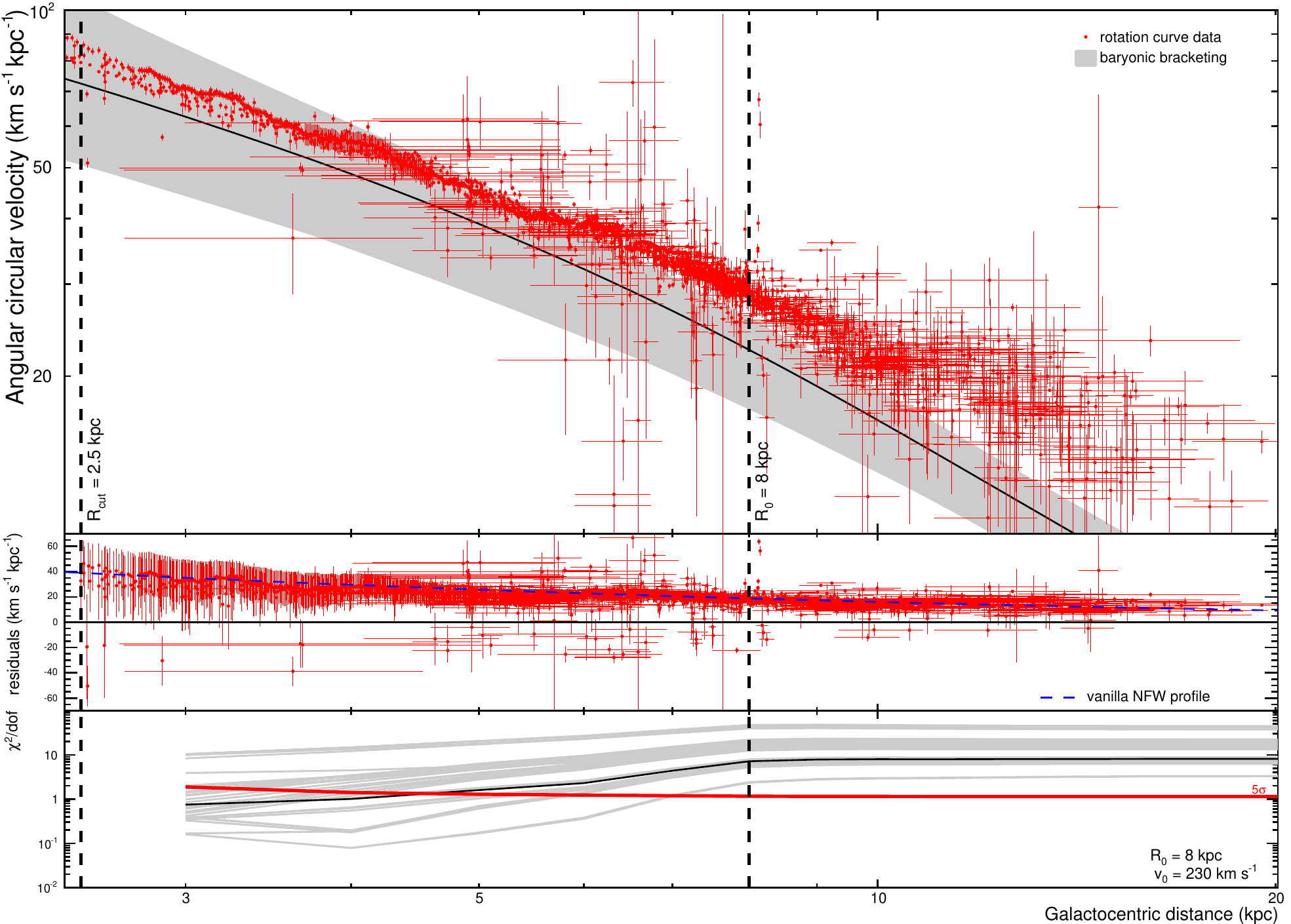}
\caption{Evidence for dark matter. In the top panel we show the angular velocity measurements from the compilation shown in Fig.~1 (red dots) together with the bracketing of the contribution of all baryonic models (grey band) as a function of galactocentric distance. Error bars correspond to 1$\sigma$ uncertainties, while the grey band shows the envelope of all baryonic models including 1$\sigma$ uncertainties. The contribution of a fiducial baryonic model is marked with the black line. The residuals between observed and predicted angular velocities for this baryonic model are shown in the middle panel. The dashed blue line shows the contribution of a Navarro-Frenk-White profile with scale radius of 20 kpc normalised to a local dark matter density of 0.4 GeV/cm$^3$. The bottom panel displays the cumulative reduced $\chi^2$ for each baryonic model as a function of galactocentric distance. The black line shows the case of the fiducial model plotted in black in the top panel, while the thick red line represents the reduced $\chi^2$ corresponding to $5\sigma$ significance. In this figure we assume a distance to the galactic centre $R_0=8$ kpc and a local circular velocity $v_0=230$ km/s, and we ignore all measurements below $R_{cut}=2.5$ kpc.}\label{fig:analysis}
\end{figure}

\clearpage

\begin{addendum}
 \item We acknowledge fruitful conversations on gas with L.~Tibaldo. The authors thank the Kavli Institute for Theoretical Physics at the University of California, Santa Barbara for hospitality during the programme ``Hunting for Dark Matter''. F.I.~thanks the K.M.~Gesellschaft for logistic support during the very early stages of this work, and the Wenner-Gren Stiftelserna for stipend support at the Oskar Klein Center in Stockholm. F.I.~also acknowledges the support of the Spanish MINECO’s ``Centro de Excelencia Severo Ochoa'' Programme under grant SEV-2012-0249 and the Consolider-Ingenio 2010 Programme under grant MultiDarkCSD2009-00064. M.~P.~acknowledges the support from Wenner-Gren Stiftelserna in Stockholm. G.B.~acknowledges the support of the European Research Council through the ERC Starting Grant {\it WIMPs Kairos}.

 \item[Competing Interests] The authors declare that they have no competing financial interests.
 \item[Correspondence] Correspondence and requests for materials should be addressed to Fabio Iocco~(email: iocco@ift.unesp.br).
\end{addendum}

\clearpage

\begin{center}
{\bf \LARGE Supplementary Information}
\end{center}

\section*{Materials and Methods}

\par $ $
\vspace{-1.5cm}
\par \underline{Rotation curve data}
\vspace{-0.25cm}
\par The main features of the compilation of rotation curve data used in this work are summarised in Table S1. Every adopted object or region has associated galactic coordinates $(\ell,b)$, heliocentric distance $d$ and heliocentric line-of-sight velocity $v_h^{los}$. The latter is usually reported in a given local standard of rest (LSR) frame rather than in the heliocentric frame, so that one has to subtract the peculiar solar motion used in the original reference to find $v_h^{los}$ and then apply the desired peculiar solar motion $\left(U,V,W\right)_{\odot}$ to get the LSR line-of-sight velocity $v_{lsr}^{los}$. We follow closely each source reference to assign errors to $d$ and $v_{lsr}^{los}$ and to account for any peculiar motion associated with specific objects.  Uncertainties on $\ell$ and $b$ are largely sub-dominant in all cases and are therefore neglected. Finally, we exclude objects with insufficient or deficient data (e.g.~on the distance determination), too close to the direction of the galactic centre or anti galactic centre and any other objects classified as suspect in the original references. After this selection, the compilation consists of 2780 individual measurements with the breakdown shown in Table S1.
\vspace{-0.5cm}
\par We then constrain the rotation curve of our Galaxy $v_c(R)$ for any given choice of the distance to the galactic centre $R_0$ and local circular velocity $v_0\equiv v_c(R_0)$. Assuming circular orbits for the objects observed (a reasonable approximation outside the influence of the galactic bulge, i.e.~$R\gtrsim R_{cut}=2.5$ kpc),
\begin{equation}\label{eqvlos}
v_{lsr}^{los} = \left( \frac{v_c(R)}{R/R_0} - v_0  \right) \cos b \, \sin \ell \quad,
\end{equation}
where $R=(d^2 \cos^2 b + R_0^2 - 2 R_0 d \cos b \, \cos \ell)^{1/2}$. For the particular case of terminal velocities, $b=0$ and $R=R_0|\sin \ell\,|$. When proper motions are available (e.g.~for open clusters and masers), similar expressions apply for the object's velocity along the longitude and latitude directions. For each object in the compilation, we have a measurement of $R$ and we invert Eq.~(\ref{eqvlos}) to obtain the angular circular velocity $w_c(R)\equiv v_c(R)/R$ and propagated error. Notice that we make use of the angular circular velocity $w_c$ rather than the actual circular velocity $v_c$, because the error of the latter is strongly correlated with the error of $R$. Instead, the errors of $w_c$ and $R$ are uncorrelated. As noticed long ago\cite{Fich1989}, using $v_c$ would introduce unnecessary complications in the statistical analysis (see below) and lead to a degradation of the accuracy.

\underline{Baryonic modelling}
\vspace{-0.25cm}
\par The exact distribution of baryons in our Galaxy is not precisely determined as of today, making it a major source of uncertainty in the present study. There are three main baryonic components: stellar bulge, stellar disk and gas. We have surveyed the literature exhaustively and collected a wide range of data-based, three-dimensional morphologies for each component. This allows for a quantitative assessment of the bracketing due to baryonic modelling, as shown in Fig.~2 in the main text. The details of bulge, disk and gas models are given below.
\vspace{-0.5cm}
\par The inner few kpc of the Milky Way are dominated by a triaxial, bar-shaped bulge of stars\cite{Dwek1995,Stanek1996,Stanek1997}. Observations clearly place the near end of the bar at positive galactic longitudes, but its precise orientation and morphology are less certain. For instance, the distribution of red clump giants in the bulge\cite{Stanek1997} is well fitted either by exponential or gaussian profiles (so-called E2 and G2 models, respectively). Apart from these two configurations, we also consider alternative truncated power-law bulges\cite{BissantzGerhard2002,Vanhollebeke2009} and a bar-shaped model with a nuclear component\cite{Zhao1996}. In view of recent developments, the possibility of an extra (long) bar\cite{LopezCorredoira2007} and a double-ellipsoid bulge\cite{Robin2012} are implemented as well. The normalisations of all seven models (and corresponding uncertainties) are fixed by matching the predicted microlensing optical depth towards $(\ell,b)=(1.50^{\circ},-2.68^{\circ})$ to the 2005 MACHO measurement $\langle \tau \rangle = 2.17^{+0.47}_{-0.38}\times 10^{-6}$\cite{MACHO2005,Iocco2011}. The microlensing contribution due to disk stars is self-consistently accounted for in accordance to the disk models described below.
\vspace{-0.5cm}
\par The stellar disk has been modelled by different authors with the help of comprehensive surveys of photometric data across the Galaxy. Typical parameterisations include thin and thick disk populations, usually featuring double-exponential profiles. We consider alternative thin plus thick configurations\cite{HanGould2003,CalchiNovatiMancini2011} as well as configurations with a stellar halo component\cite{deJong2010,Juric2008}. The single maximal disk suggested recently\cite{BovyRix2013} is also implemented. All five models are normalised to the latest local surface density constraint $\Sigma_{*}=38\pm 4\textrm{ M}_\odot\textrm{/pc}^2$\cite{BovyRix2013}, from which we propagate the uncertainty to the disk component.
\vspace{-0.5cm}
\par Finally, a non-negligible part of the baryons in the Milky Way is in the form of gas, namely molecular, atomic and ionised hydrogen and heavier elements. The distribution of each component is relatively well-known but extremely irregular. This is for example the case of the gas within 10 pc of the galactic centre\cite{Ferriere2012}, which for our purposes can be safely considered as a point-like mass. For the inner 2 kpc, we model molecular and atomic hydrogen in the central molecular zone and holed disk, and the distribution of ionised hydrogen is split into its warm, hot and very hot phases\cite{Ferriere2007}. In the range $R=2-20$ kpc, instead, two alternative morphologies\cite{Ferriere1998,Moskalenko2002} are used for each gas component. We set up in this way our two gas models, whose uncertainties are assigned by taking a CO-to-H$_2$ conversion factor of $(2.5-10)\times 10^{19}\textrm{ cm}^{-2}\textrm{(K km/s)}^{-1}$ for $R<2$ kpc and $(0.5-3.0)\times 10^{20}\textrm{ cm}^{-2}\textrm{(K km/s)}^{-1}$ for $R>2$ kpc\cite{Ferriere1998,Ackermann2012}.
\vspace{-0.5cm}
\par Once a model for bulge, disk and gas is specified, the individual gravitational potentials (and thus the individual contributions to the rotation curve) can be easily computed through multipole expansion\cite{BinneyTremaine}. Expanding up to $l_{max}=2$ (see below for a convergence test) and averaging over the azimuthal direction, we can then derive the overall baryonic contribution to the rotation curve $\omega_b^2=\omega_{bulge}^2+\omega_{disk}^2+\omega_{gas}^2$ and the corresponding propagated uncertainties.

\underline{Statistical analysis}
\vspace{-0.25cm}
\par The central task in this work is to compare the observed rotation curve of the Galaxy $\omega_c(R)$ to that expected from baryons $\omega_b(R)$, and decide whether $\omega_c^2-\omega_b^2$ is compatible with zero or not. This task is complicated by the sizeable error on $R$ in the rotation curve data, especially at intermediate and large $R$ as shown in Fig.~1 of the main text. Given the large amount of observations (see Table S1), a customary technique usually adopted in the literature is that of binning the data. Since binning entails loss of information, we opt not to do it and use instead the full power of the data taking proper account of galactocentric distance errors. (We did however check explicitly that a binned analysis with a weighted mean of the measurements just reinforces the results presented in the main text.) Following Ref.~\cite{Fich1989} and introducing the reduced variables $x=R/R_0$ and $y=\omega/\omega_0-1$ (with $\omega_0=v_0/R_0$), the two-dimensional $\chi^2$ reads
\begin{equation}\label{chi2}
\chi^2 = \sum_{i=1}^{N} { d_i^2 } \equiv \sum_{i=1}^{N} { \left[ \frac{(y_i-y_{b,i})^2}{\sigma_{y,i}^2} + \frac{(x_i-x_{b,i})^2}{\sigma_{x,i}^2}  \right] } \quad,
\end{equation}
where $(x_{b,i},y_{b,i})$ is the point in the baryonic curve $y_b(x)=\omega_b(R=xR_0)/\omega_0-1$ that minimises $d_i$. Notice that the expression above can be applied because the errors on $R$ and $\omega$ (i.e.~$x$ and $y$) are uncorrelated and hence the error ellipse is not tilted in the $(R,\omega)$ plane. This would not be the case for the error-correlated pair $(R,v_c)$. We have performed Monte Carlo calculations for a fiducial baryonic model and typical uncertainties $\sigma_{x,i},\sigma_{y,i}$, and verified that the statistic in Eq.~(\ref{chi2}) follows approximately a $\chi^2$ distribution. The lower panel in Fig.~2 in the main text shows the reduced chi-square $\chi^2/N$, where the sum in Eq.~(\ref{chi2}) is restricted to objects with galactocentric distances below a given $R$. Due to the breakdown of the assumption of circular orbits, all objects with $R\leq R_{cut}=2.5$ kpc are ignored in the analysis.

\underline{Robustness of the results}
\vspace{-0.25cm}
\par Our main findings are presented in Fig.~2 for a wide bracketing of baryonic models. This illustrates already the robustness of the results. However, Fig.~2 was obtained for fixed (albeit reasonable) choices of galactic parameters and data selection. Here we show that different choices lead to the same conclusion as in the main text, i.e.~that baryons cannot explain alone the observed rotation curve in the inner Galaxy.
\vspace{-0.5cm}
\par We start by varying the fundamental galactic parameters, namely $R_0$, $v_0$ and $\left(U,V,W\right)_{\odot}$. The existing determinations of $R_0$ (e.g.~\cite{Gillessen2009,Ando2011,Malkin2012,Reid2014}) do not all agree with each other, but are confined to a reasonably narrow range, $8.0\pm 0.25$ kpc\cite{Malkin2012}. We adopt a more conservative range $8.0\pm 0.5$ kpc in our tests. A similar situation holds for $v_0$ (e.g.~\cite{Reid2009,Bovy2009,McMillan2010,Bovy2012,Reid2014}) where the range $230\pm 20$ km/s encompasses most determinations. As for the peculiar solar motion, we take $\left(U,V,W\right)_{\odot}=(11.10,12.24,7.25)$ km/s\cite{Schoenrich2010} as our fiducial values, but consider as well other determinations in the literature\cite{DehnenBinney1998,Bovy2012,Reid2014}: $\left(U,V,W\right)_{\odot}=(10.00,5.25,7.17)$ km/s, $\left(U,V,W\right)_{\odot}=(10.00,26.00,7.25)$ km/s (with $v_0=218$ km/s) and $\left(U,V,W\right)_{\odot}=(10.7,15.6,8.9)$ km/s (with $v_0=240$ km/s). The effect of taking different choices of galactic parameters is shown in Fig.~S1. Also shown is the evidence obtained when using gas kinematics, star kinematics and masers separately. Furthermore, the convergence of the multipole expansion used to compute $\omega_b$ was tested up to $\ell_{max}=8$. Finally, we checked the impact of streaming motions due to spiral arms by adding a 11.8 km/s systematic\cite{BrandBlitz1993} to each rotation curve measurement. In all cases the discrepancy between observed and predicted rotation curves is high ($>5\sigma$) and very robust against galactic nuisances, data selection and systematics.

\renewcommand{\thefigure}{S\arabic{figure}} 
\setcounter{figure}{0}

\begin{figure}

\includegraphics[width=0.32\textwidth,height=0.3\textwidth]{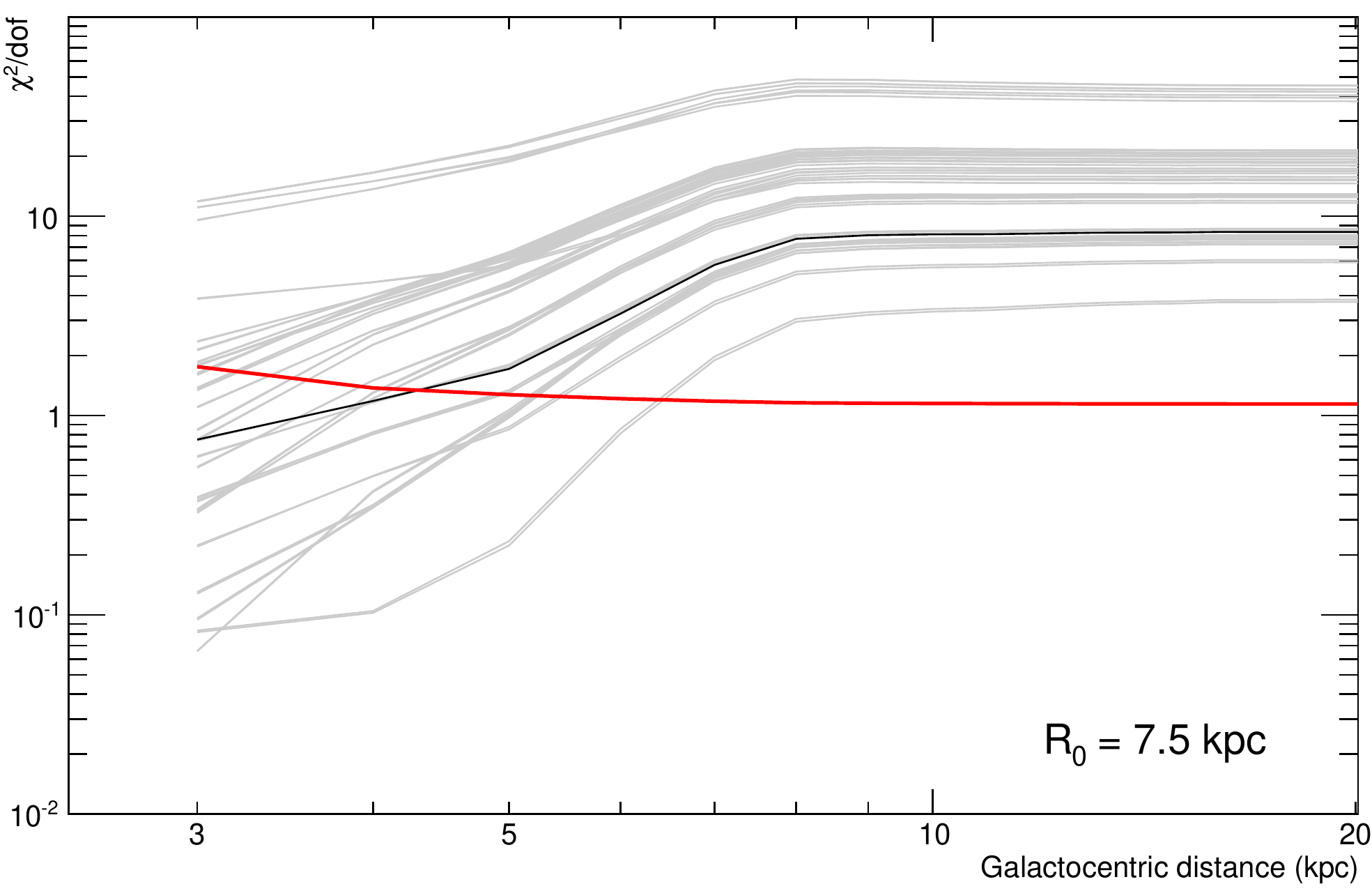}
\includegraphics[width=0.32\textwidth,height=0.3\textwidth]{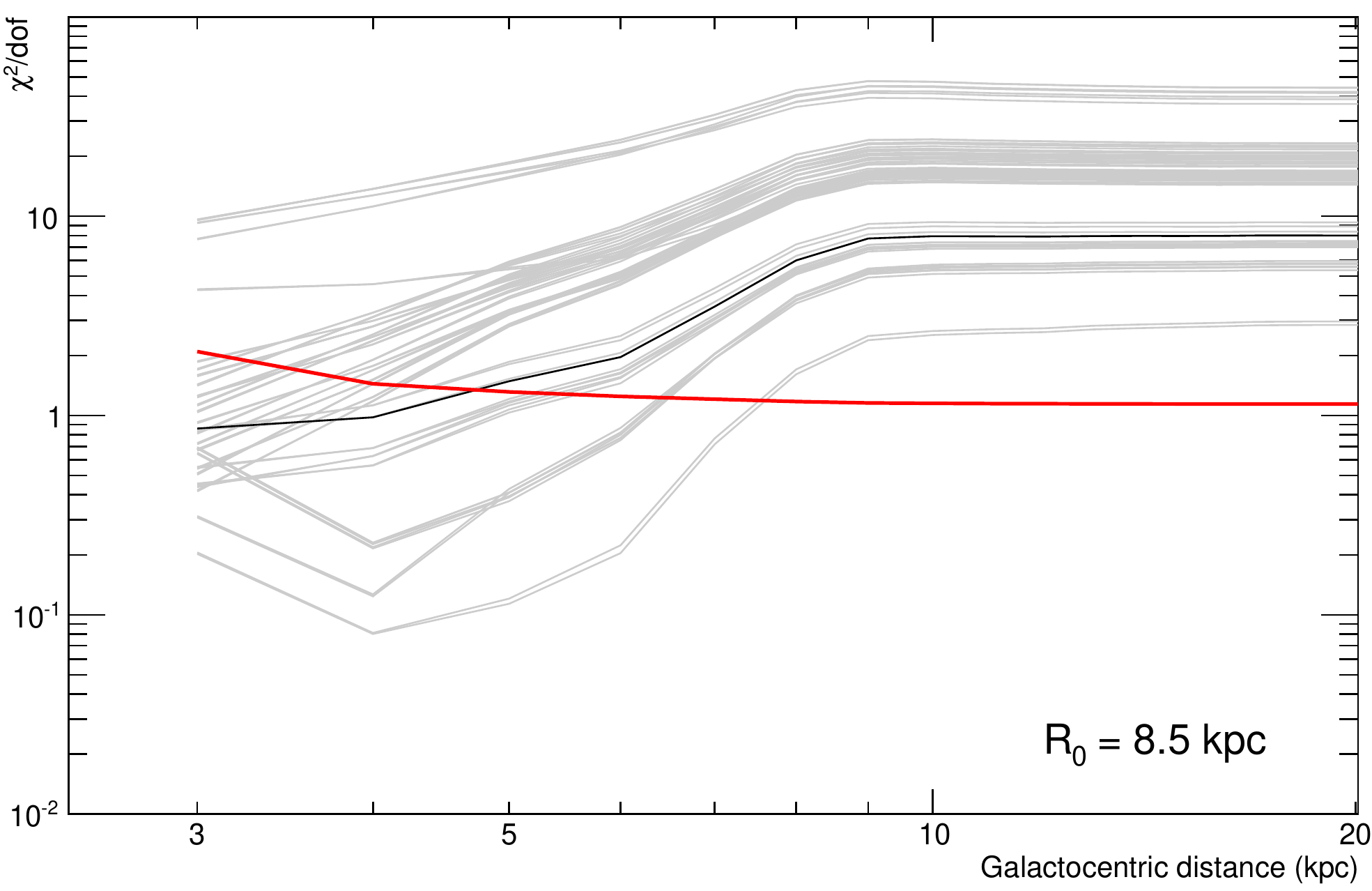}
\includegraphics[width=0.32\textwidth,height=0.3\textwidth]{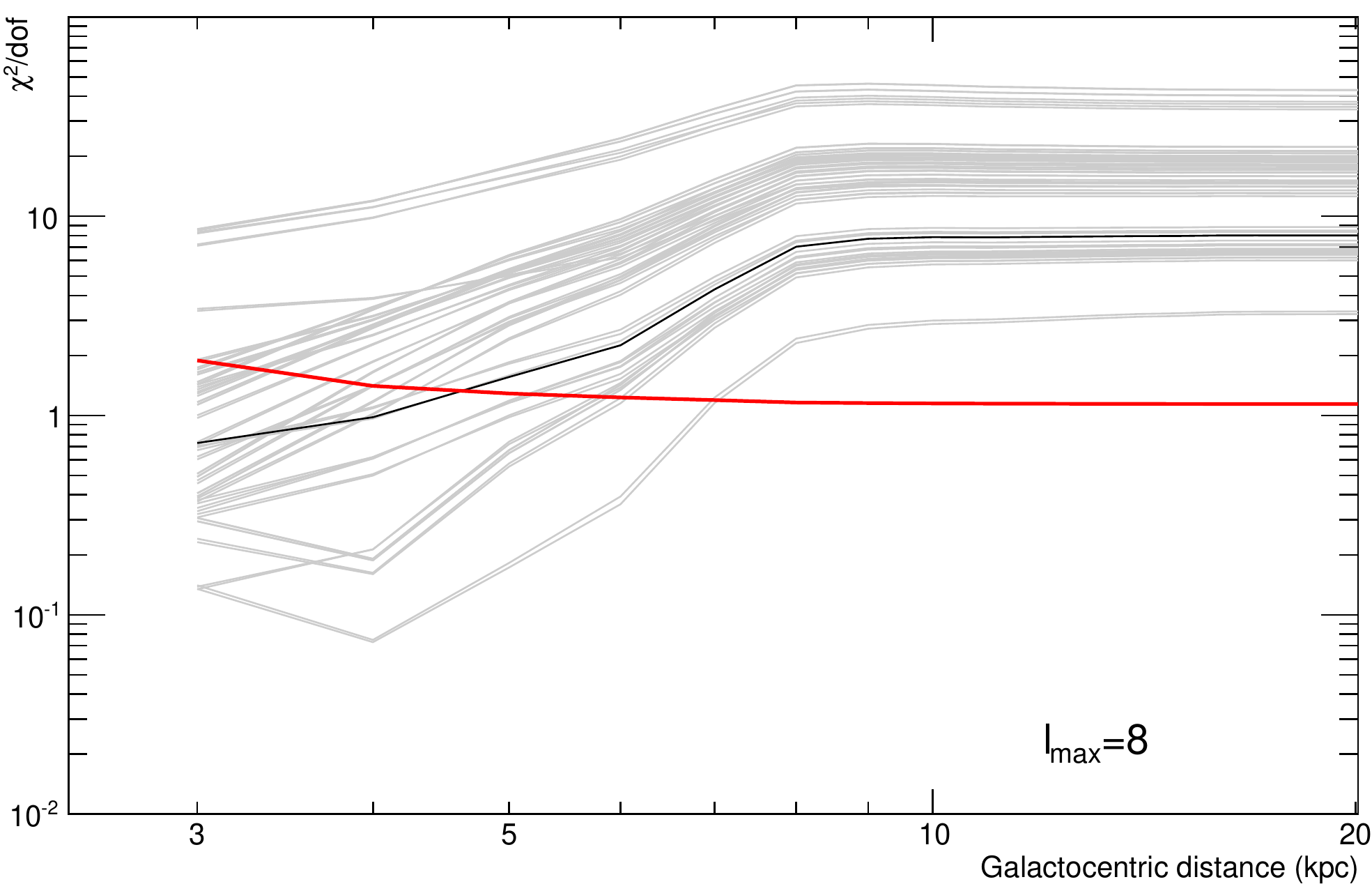}

\includegraphics[width=0.32\textwidth,height=0.3\textwidth]{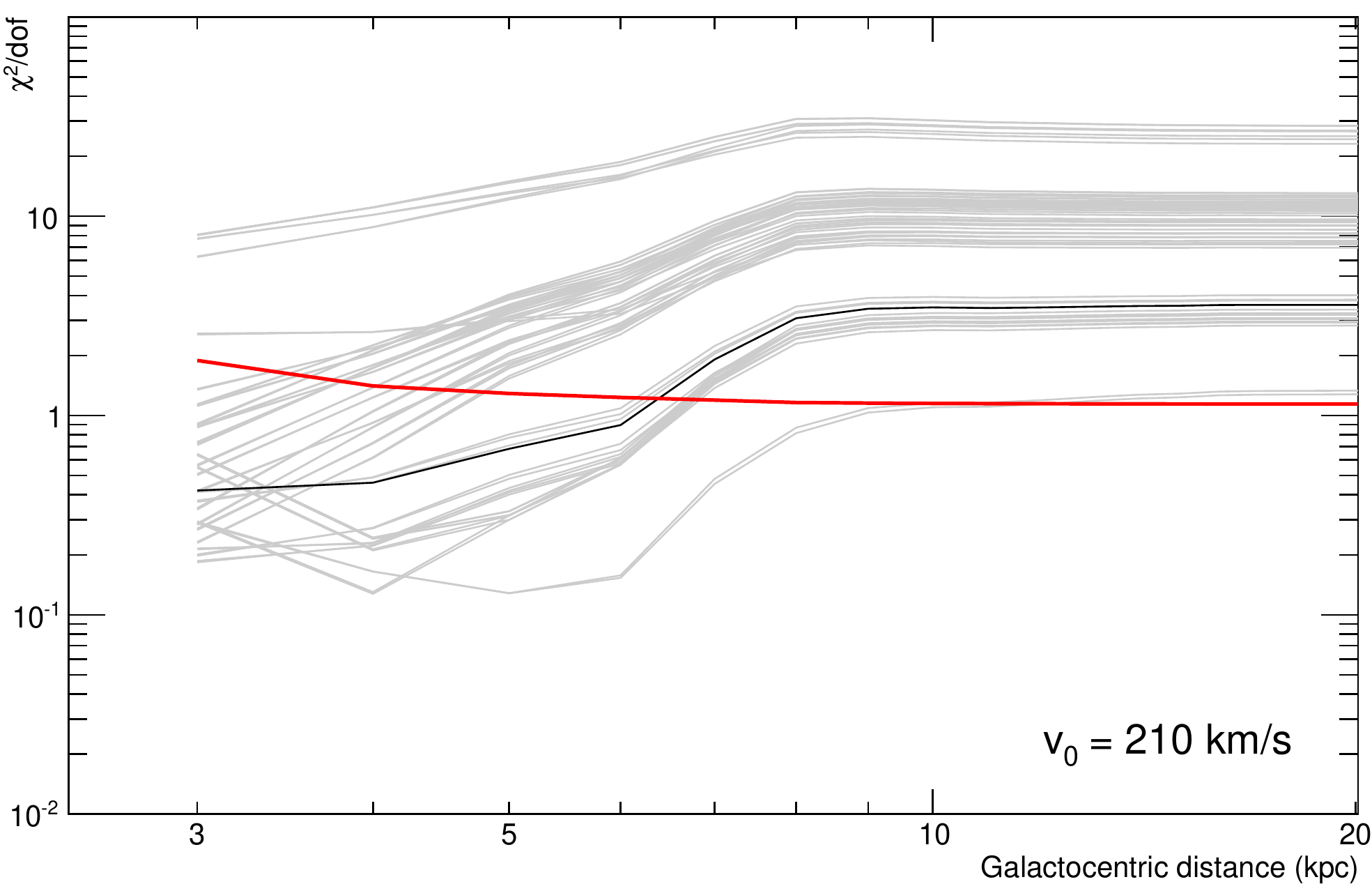}
\includegraphics[width=0.32\textwidth,height=0.3\textwidth]{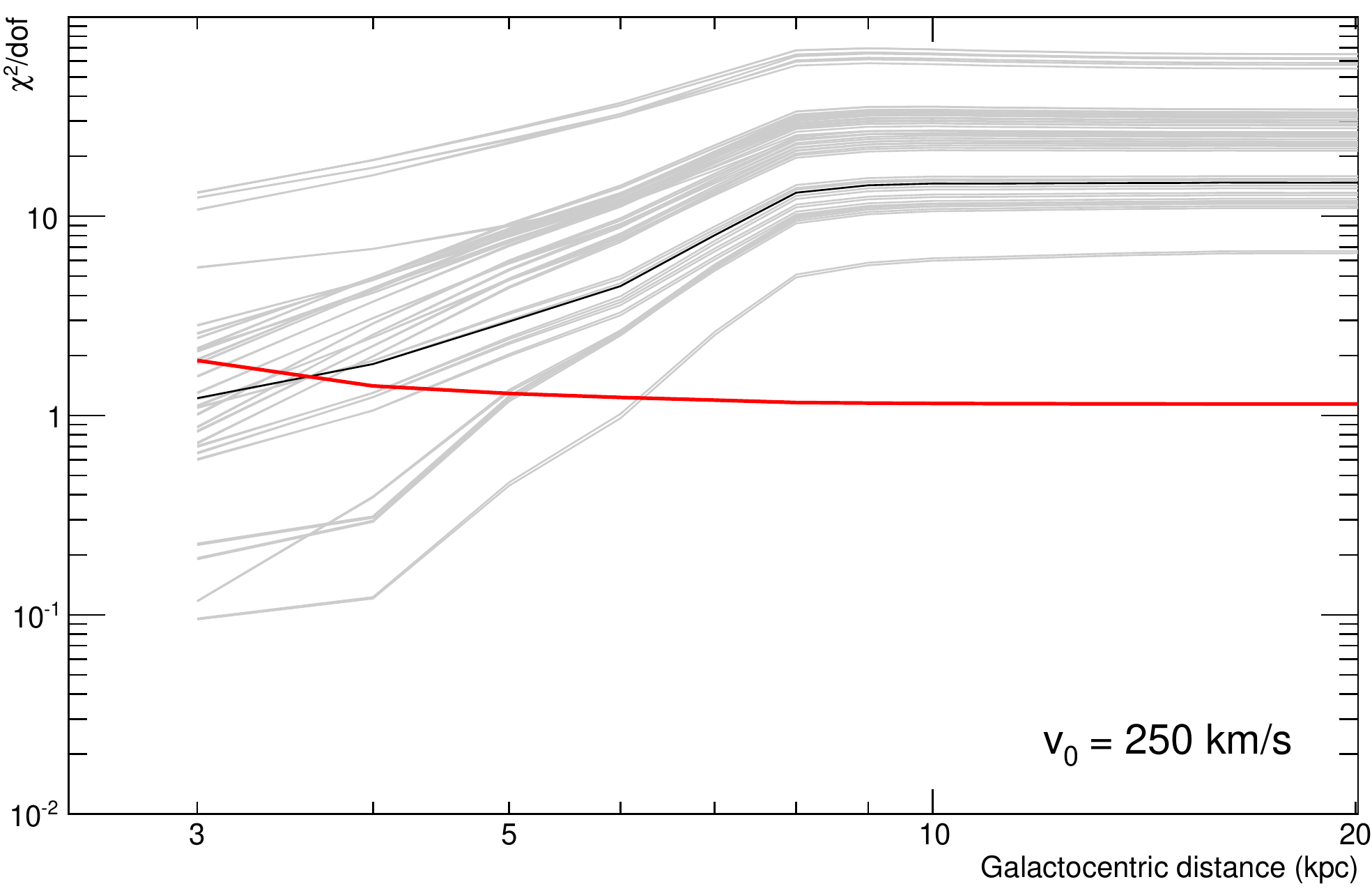}
\includegraphics[width=0.32\textwidth,height=0.3\textwidth]{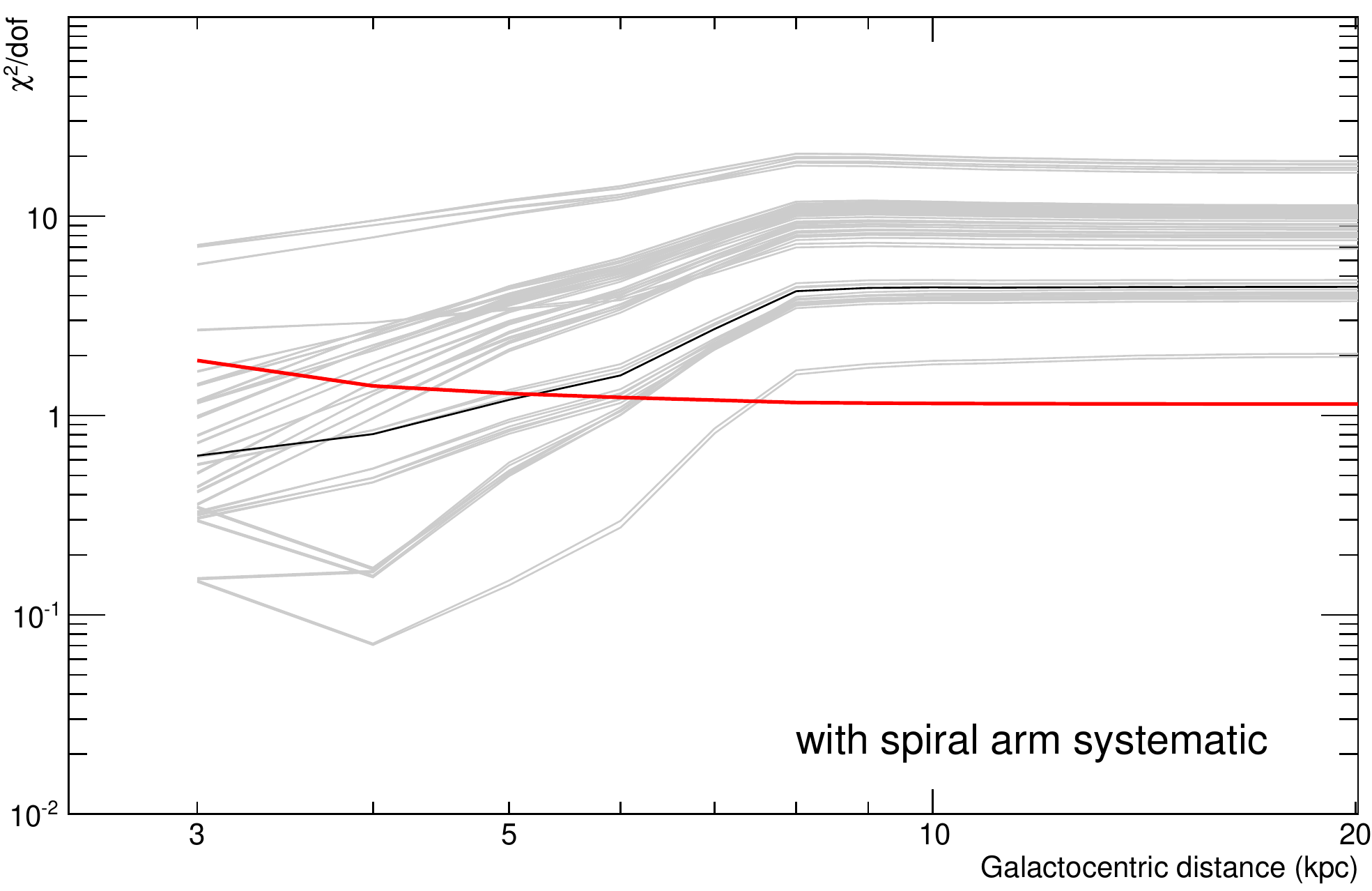}

\includegraphics[width=0.32\textwidth,height=0.3\textwidth]{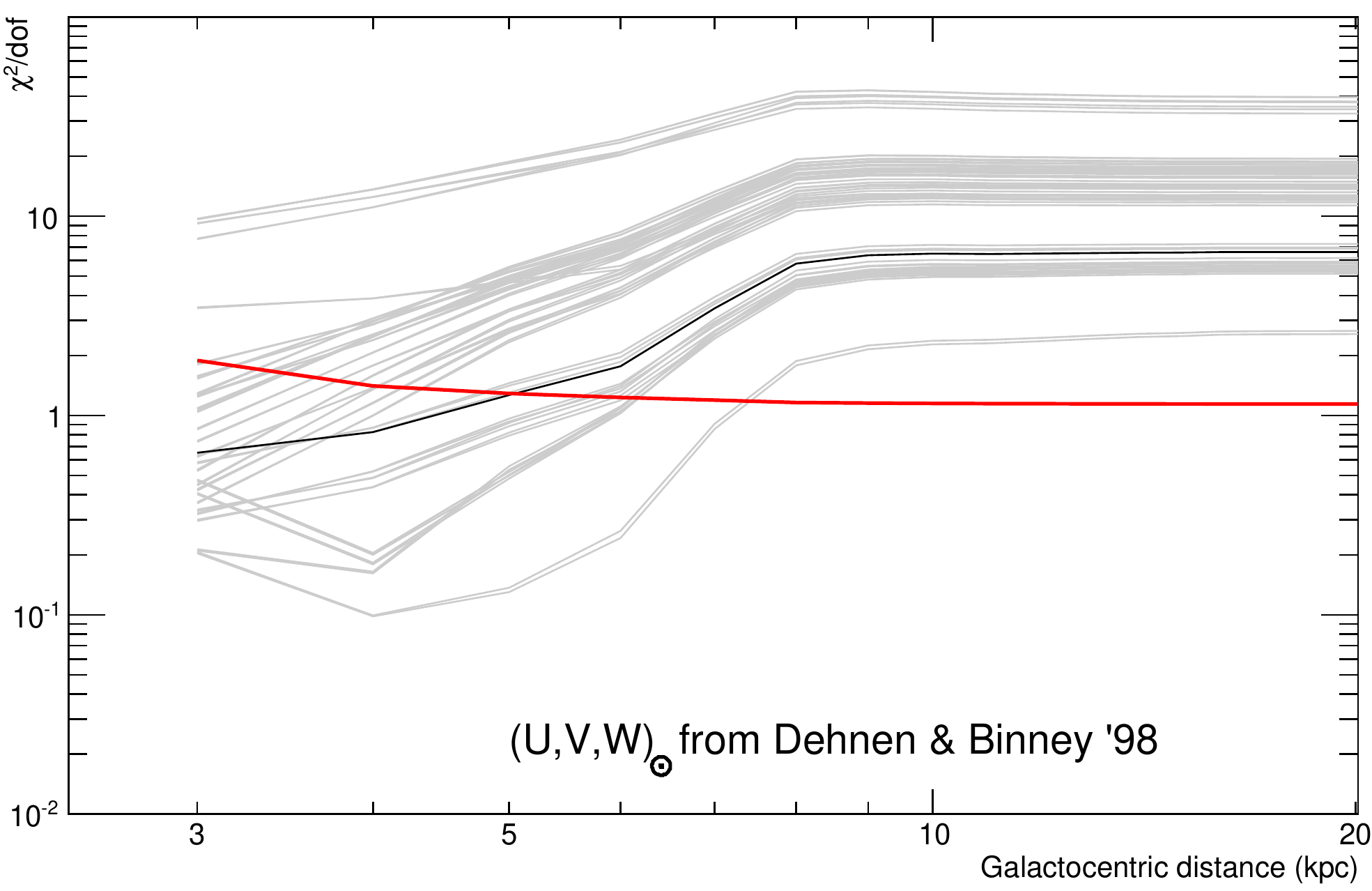}
\includegraphics[width=0.32\textwidth,height=0.3\textwidth]{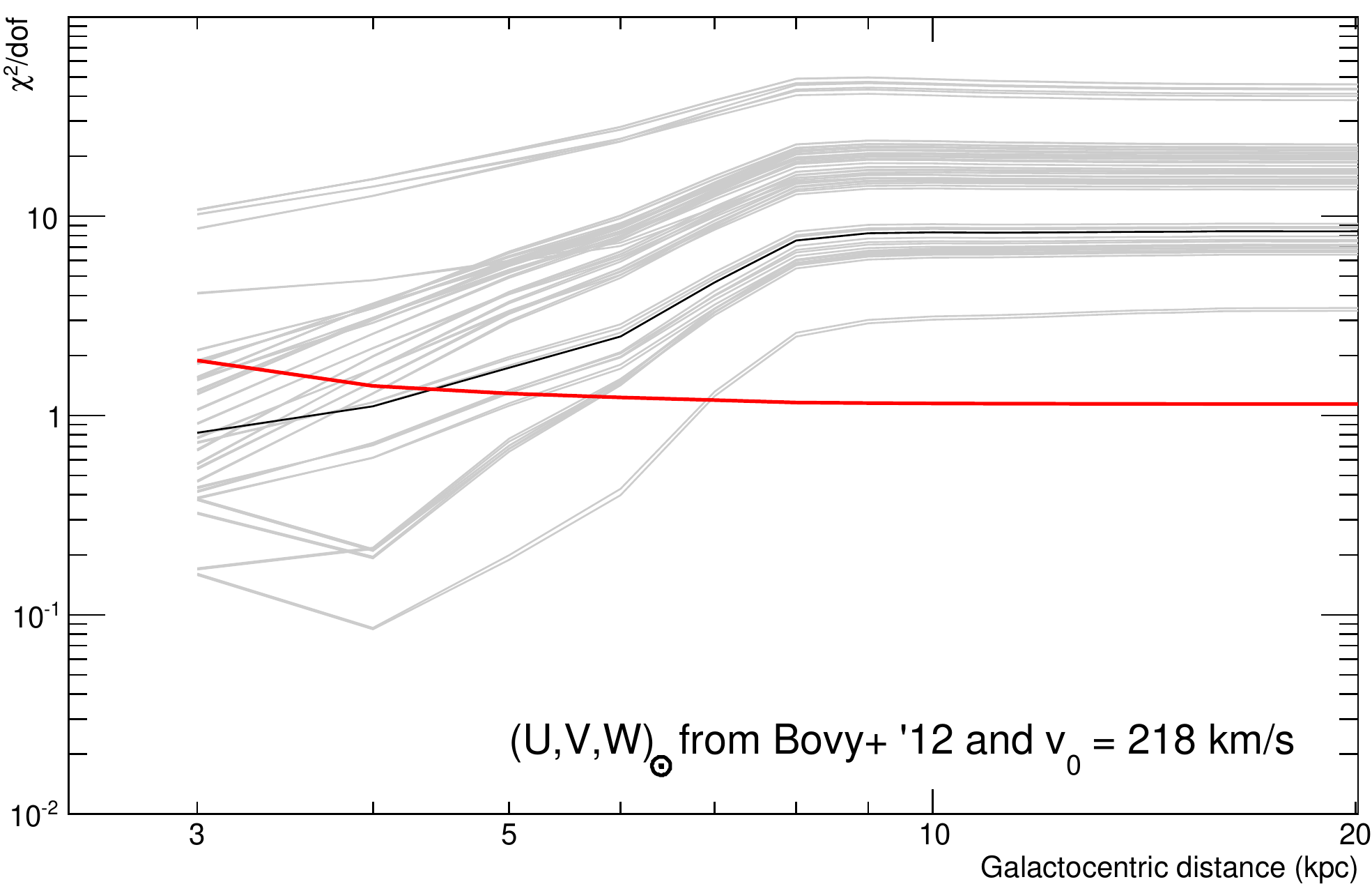}
\includegraphics[width=0.32\textwidth,height=0.3\textwidth]{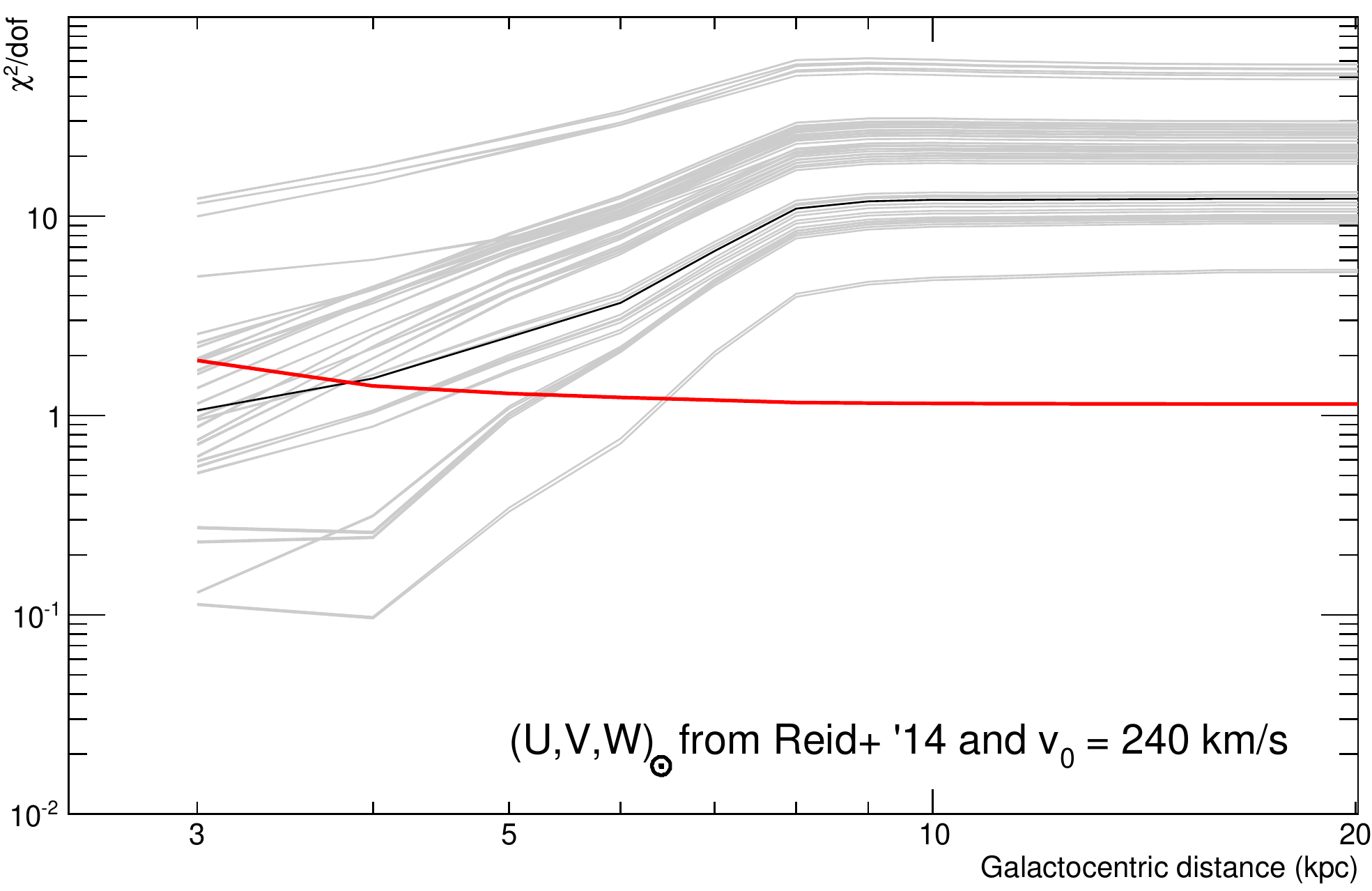}

\includegraphics[width=0.32\textwidth,height=0.3\textwidth]{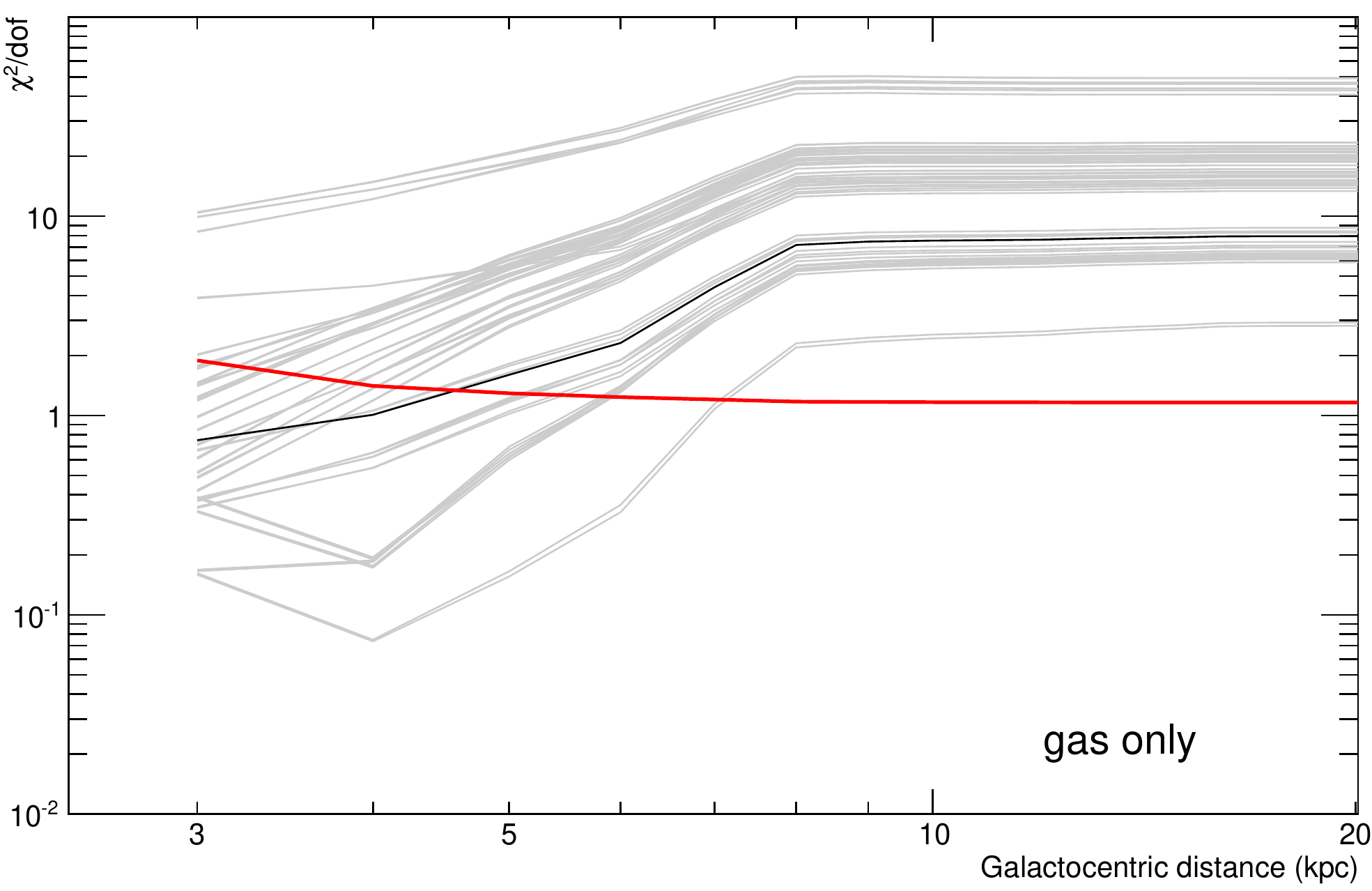}
\includegraphics[width=0.32\textwidth,height=0.3\textwidth]{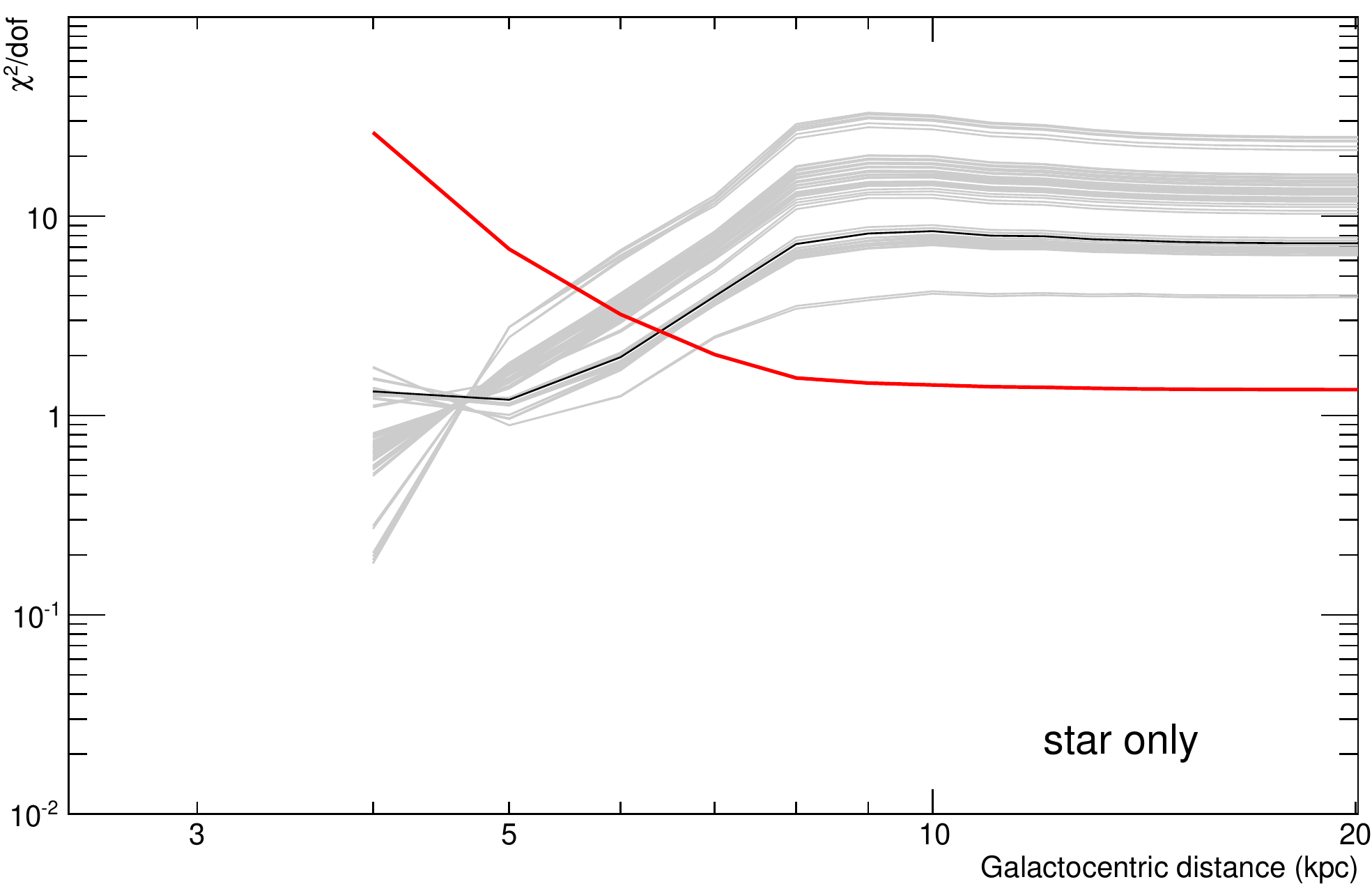}
\includegraphics[width=0.32\textwidth,height=0.3\textwidth]{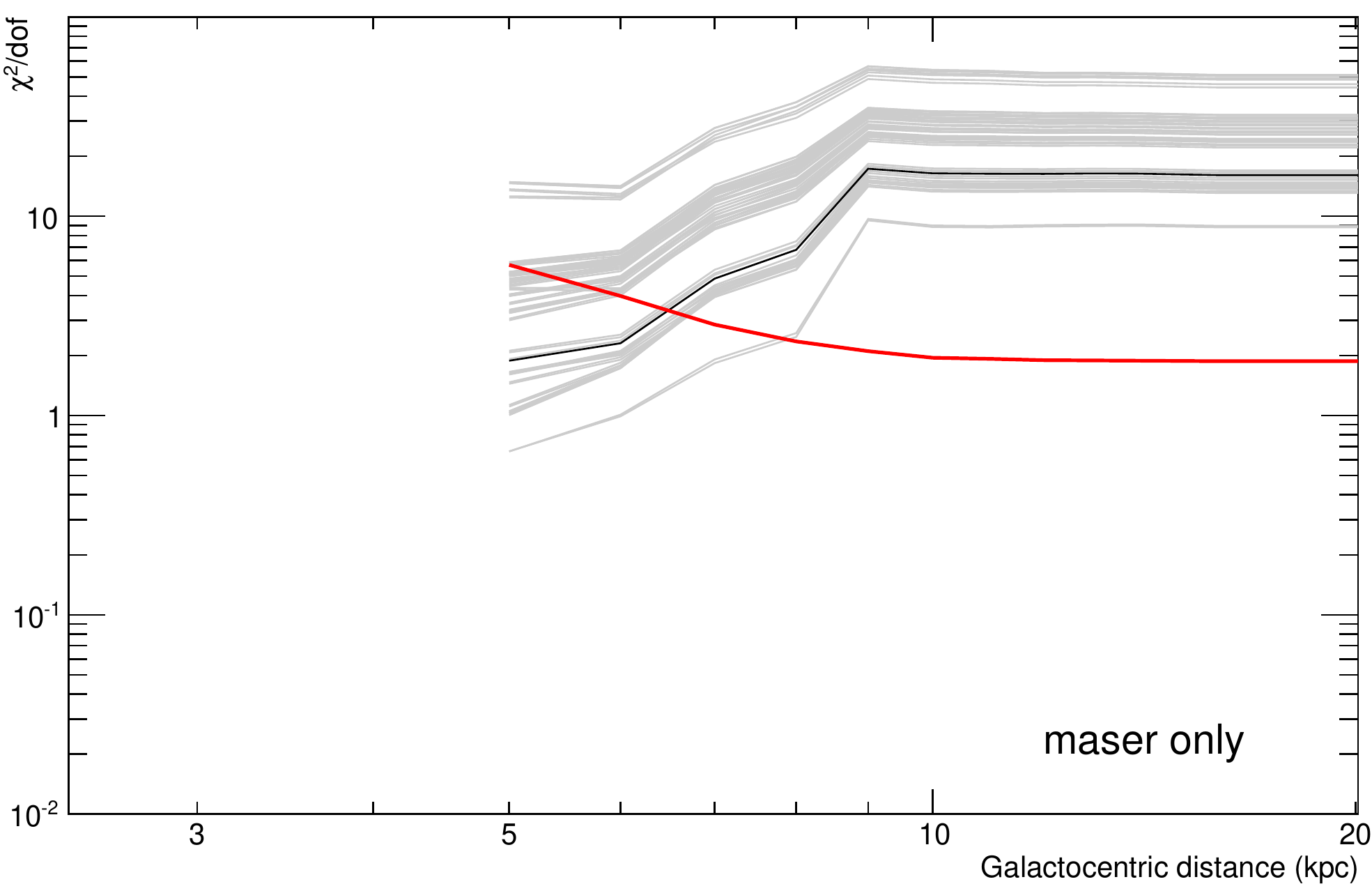}

\caption{The evidence for dark matter in the inner Galaxy and its dependence on galactic parameters, data selection and systematics. The plots show the cumulative reduced $\chi^2$ for each baryonic model as a function of galactocentric distance. The thick red line represents the reduced $\chi^2$ corresponding to $5\sigma$ significance, while the black line shows the case of the fiducial baryonic model.}
\end{figure}


\renewcommand{\thetable}{S\arabic{table}} 
\setcounter{table}{0}

\begin{table}
\begin{center}
\small
\begin{tabular}{l c  c c c} 
{\bf Object type} 		       						& {\bf $\boldsymbol{R}$ [kpc]}			& {\bf quadrants} 				& {\bf \# objects}	\\
\hline
HI terminal velocities 								&						& 						&	 		\\
$\quad$ Fich+ '89\cite{Fich1989} 						& \hspace{0.21cm}2.1 -- 8.0\hspace{0.21cm}	& 1,4						& 149	 		\\
$\quad$ Malhotra '95\cite{Malhotra1995} 					& \hspace{0.21cm}2.1 -- 7.5\hspace{0.21cm} 	& 1,4		 				& 110  			\\
$\quad$ McClure-Griffiths \& Dickey '07\cite{McClure-GriffithsDickey2007} 	& \hspace{0.21cm}2.8 -- 7.6\hspace{0.21cm} 	& 4		 				& 701 			\\
HI thickness method								&						&						&			\\
$\quad$ Honma \& Sofue '97\cite{HonmaSofue1997} 				& \hspace{0.21cm}6.8 -- 20.2 			& --		 				& \hspace{0.21cm}13 	\\
CO terminal velocities								&						& 						&			\\
$\quad$ Burton \& Gordon '78\cite{BurtonGordon1978}  				& \hspace{0.21cm}1.4 -- 7.9\hspace{0.21cm} 	& 1						& 284  			\\
$\quad$ Clemens '85\cite{Clemens1985}  					& \hspace{0.21cm}1.9 -- 8.0\hspace{0.21cm}	& 1		   					& 143  			\\
$\quad$ Knapp+ '85\cite{Knapp1985}  						& \hspace{0.21cm}0.6 -- 7.8\hspace{0.21cm} 	& 1		 				& \hspace{0.21cm}37 	\\
$\quad$ Luna+ '06\cite{Luna2006} 						& \hspace{0.21cm}2.0 -- 8.0\hspace{0.21cm} 	& 4						& 272  			\\
HII regions									&						&						&			\\
$\quad$ Blitz '79\cite{Blitz1979} 						& \hspace{0.21cm}8.7 -- 11.0 			& 2,3						& \hspace{0.42cm}3 	\\
$\quad$ Fich+ '89\cite{Fich1989}  						& \hspace{0.21cm}9.4 -- 12.5 			& 3						& \hspace{0.42cm}5  	\\
$\quad$ Turbide \& Moffat '93\cite{TurbideMoffat1993}   			& 11.8 -- 14.7 					& 3						& \hspace{0.42cm}5 	\\
$\quad$ Brand \& Blitz '93\cite{BrandBlitz1993} 				& \hspace{0.21cm}5.2 -- 16.5 			& 1,2,3,4					& 148			\\
$\quad$ Hou+ '09\cite{Hou2009}  						& \hspace{0.21cm}3.5 -- 15.5 			& 1,2,3,4					& 274			\\
giant molecular clouds								&						&						&			\\
$\quad$ Hou+ '09\cite{Hou2009}							& \hspace{0.21cm}6.0 -- 13.7 			& 1,2,3,4					& \hspace{0.21cm}30  	\\
\hline
open clusters									&						&						&			\\
$\quad$ Frinchaboy \& Majewski '08\cite{FrinchaboyMajewski2008}  		& \hspace{0.21cm}4.6 -- 10.7 			& 1,2,3,4					& \hspace{0.21cm}60 	\\
planetary nebulae								&						& 						&			\\
$\quad$ Durand+ '98\cite{Durand1998} 						& \hspace{0.21cm}3.6 -- 12.6 			& 1,2,3,4					& \hspace{0.21cm}79 	\\
classical cepheids								&						&						&			\\
$\quad$ Pont+ '94\cite{Pont1994} 						& \hspace{0.21cm}5.1 -- 14.4 			& 1,2,3,4					& 245			\\
$\quad$ Pont+ '97\cite{Pont1997} 						& 10.2 -- 18.5 					& 2,3,4						& \hspace{0.21cm}32 	\\
carbon stars									&						&						&			\\
$\quad$ Demers \& Battinelli '07\cite{DemersBattinelli2007} 			& \hspace{0.21cm}9.3 -- 22.2 			& 1,2,3						& \hspace{0.21cm}55 	\\
$\quad$ Battinelli+ '13\cite{Battinelli2013} 					& 12.1 -- 24.8 					& 1,2						& \hspace{0.21cm}35 	\\
\hline
masers										&						&						&			\\
$\quad$ Reid+ '14\cite{Reid2014}  						& \hspace{0.21cm}4.0 -- 15.6 			& 1,2,3,4					& \hspace{0.21cm}80	\\
$\quad$ Honma+ '12\cite{Honma2012} 						& \hspace{0.21cm}7.7 -- 9.9\hspace{0.21cm} 	& 1,2,3,4					& \hspace{0.21cm}11 	\\
$\quad$ Stepanishchev \& Bobylev '11\cite{StepanishchevBobylev2011} 		& 8.3						& 3						& \hspace{0.42cm}1 	\\
$\quad$ Xu+ '13\cite{Xu2013} 							& 7.9 						& 4		 				& \hspace{0.42cm}1 	\\
$\quad$ Bobylev \& Bajkova '13\cite{BobylevBajkova2013} 			& \hspace{0.21cm}4.7 -- 9.4\hspace{0.21cm} 	& 1,2,4						& \hspace{0.42cm}7 	\\
\hline
\end{tabular}
\caption{Our compilation of rotation curve data for the Milky Way. For each object type and reference, we report the range of galactocentric distance $R$ (assuming a distance to the galactic centre $R_0=8$ kpc), the galactic quadrant(s) as well as the number of objects analysed (after cuts).}
\end{center}
\end{table}

\end{document}